# Interpretation as a factor in understanding flawed spreadsheets

David A. Banks and Ann Monday
University of South Australia, Adelaide, Australia
Email {David.Banks}, {Ann.Monday}@unisa.edu.au

**ABSTRACT**

*The spreadsheet has been used by the business community for many years and yet still raises a number of significant concerns. As educators our concern is to try to develop the students skills in both the development of spreadsheets and in taking a critical view of their potential defects. In this paper we consider both the problems of mechanical production and the problems of translation of problem to spreadsheet representation.*

## 1. INTRODUCTION

The spreadsheet has been used as a decision support tool for a considerable number of years but still suffers from significant defects in terms of the way that they 'inform', or arguably 'misinform', decision makers. Spreadsheets, at least in the hands of the typical end user, have proved to be less effective as decision support tools than may have been expected. Although they apparently offer decision makers the ability to develop models of business problems and to test various possibilities by manipulating the data in the model, two key problem areas emerge in practice. These areas relate to errors made in the mechanical production of the spreadsheet and also to errors in the translation of the problem situation into a model that is then represented as a spreadsheet. These two causes of error can lead to a position where decision makers may act in the belief that decisions can be made with confidence on the output from the spreadsheet despite evidence to the contrary. In this paper we consider both the problems of mechanical production and the problems of translation of problem to spreadsheet representation, with a strong interpretive focus on the latter. Our approach is therefore biased towards the 'softer', interpretive dimensions of the information systems field and views the spreadsheet as being a tool located in a socio-technical system. We do, however, also draw upon some of the 'hard' programming literature as this provides useful insight to some aspects of the spreadsheet development process. We develop our considerations primarily within the context of the teaching of subjects that utilise spreadsheets as decision support tools and illustrate our arguments with examples of student work.

## 2. PROBLEMS IN SPREADSHEET DEVELOPMENT

Since the proliferation of PCs in the early 80s businesses have seen a continuing growth in the use of applications created by end-users for themselves and others in their organisations. Frenzel (1992) argues that User Developed Applications (UDAs) have grown because they offer greater user control and increased flexibility and responsiveness for the user. They encourage innovation, reduce the workload of the IT department and enable the user to achieve what they want in a much shorter time-scale. Thus they allow changes to be implemented quickly and the effects to be seen immediately. McGill (2000) notes that UDAs now form a significant proportion of organisational information systems. The downside of this faster and user directed development has, however, led to the emergence of major problems that can have significant and negative impact on businesses that rely heavily on spreadsheets.

Alavi and Weiss (1985) note that 'In the enthusiasm to benefit from EUC activities, corporations are overlooking the potential risks of these activities. Organizational exposure to EUC is costly.' They point to problems of lack of time devoted to problem definition, lack of specification, solving the wrong problem, selection of inappropriate software tool, lack of documentation, lack of testing and validation etc. The effect of poor development, testing and maintenance processes can typically impact upon bottom line values, tax payments, negative balance for stock on hand and significant underestimates in project costings (Panko and Halverson, 1996). Panko (2000) cites a variety of studies that indicate a high percentage of all spreadsheets contain errors. Audit work carried out by consulting organisations have found that up to 90% of all spreadsheets with more than 150 rows contained errors. Many of these spreadsheets were of high importance to the organisations concerned. One survey of 106 spreadsheets indicated that 49% of the surveyed spreadsheets created new corporate data and only 7% of spreadsheets were considered to be of 'low' importance (Hall, M.J.J. 1996, cited in Panko and Halverson 1996).

Other research examined the risks associated with spreadsheet errors. Teo and Tan (1999) found that '…spreadsheet applications were used by an overwhelming 91% of a sample of end-users', whilst Janvrin and Morrison (2000) comment that '...as many as 44% of all end-user spreadsheets contain at least one error'. Since 1987 researchers have shown that the problem of spreadsheet errors in UDAs is increasing. Teo and Tan (1999) state that 'This growing concern over spreadsheet errors can be attributed to the increasing popularity of spreadsheets to support important financial analysis, budgeting, and forecasting applications.' They have also seen '...a tendency for end-users to view spreadsheets as simple tools and to be overconfident about the error-free nature of their spreadsheets.' McGill (2000) noted UDAs that may be incorrect in design, inadequately tested, and poorly maintained. Teo and Tan (1999) also found that 'inadequate care is taken to design spreadsheets, which in turn makes error detection and correction even more difficult' and the quality of the application is reduced further.

Edberg and Bowman (1996) recognised that 'UDAs represent a considerable risk to organizations since users who create applications frequently have little or no training in development methods'. The need for end-users to be given some design and implementation training is highlighted by Hobby (1996). Ross and Ruhleder (1993), in exploring the range of skills required by IS professionals suggest that IS educators should impart not only the technical skills to students but also provide them with an awareness of a wide range of technical, social and organisational concerns.

Further research into spreadsheets has included Chan and Storey (1996) who investigated the use of spreadsheets within the organisation to determine the relationship between usage, tasks, proficiency and satisfaction. This research determined 'proficiency has a greater impact on the tasks than the tasks have on proficiency; users do not often use the commonly available spreadsheet features; users prefer software packages they know and understand rather than the best package for the task; user proficiency was not related to the importance of the decision made as a result of the spreadsheet analysis'.

There are a number of techniques and tools that can be used to verify at least the mathematical or formulaic integrity of spreadsheets, based on their quantitative nature. The causal relationships of the data in spreadsheets can be rigorously investigated and tests can be applied to check that with given input data the process modelled in the spreadsheet will provide an accurate output. It is thus possible to check that the data is correct and is in the required format, that the relationship of the data is logical, formulae are valid and so on. These can be viewed as representing the 'hard' aspects of the spreadsheet and errors here may be attributed to typing errors, incorrect linking of cells, incorrect use of formulae and a variety of other technical causes. However, our work with student spreadsheets indicates that while these hard errors do occur the main problem lies in the translation, or interpretation, of a given problem into a spreadsheet representation of that problem. Our purpose in using spreadsheets is to develop their understanding of a spreadsheet as not just a quantitative, number processing tool, but also as a qualitative tool that informs the decision maker rather than providing the answer.

Our experience with students over a number of years shows they continue to make such common errors in spreadsheets as identified in taxonomies of errors (Rajalingham, Chadwick and Knight 2000 and Panko 2000) - including poor cell protection (often none at all), poor use of formulae (including embedded numbers), lookup poorly used, no validation, clear inconsistencies in data output that should have been checked and assumptions not clearly stated.

We also consistently see attempts to compare 'apples with oranges' and incorrect conversions of data into common format (litres/gallons, kilograms/tonnes, dollars/yen). Had these students been developing the spreadsheet in a real business setting it is evident that the results would have had little real value for a decision maker. They made an assumption that there was a single 'right' answer, and that their model generated this 'right' answer. This was despite the fact that they were told the spreadsheet would be used as a decision *support* aid for a group of managers and would, in all probability, not provide a single definitive answer. There is a tendency to adopt a view of spreadsheet development that echoes Rushby's (1980) comments about computer program development: 'It is tempting, after the white heat of coding, to assume that if the program compiles and runs without any execution errors, then it works. Perhaps it does work, but all that proves is that the program does something – not that it does the right thing. Only thorough testing can we evaluate the program, and demonstrate that it does indeed meet its' specification'

We are not suggesting that students are incapable of developing good spreadsheets. Our courses are not training courses and seek to expose students to the difficulties of developing decision support models using a tool that itself has a number of limitations. Klein and Methlie (1995) suggest that the problems with spreadsheets in the context of decision support systems include:

- "they do not provide a clear understanding of an application global logic in using resources such as data, decision models, reports, forms, etc;
- they do not provide satisfactory readability of decision models;
- they do not provide a way to represent easily data structures more complex than two-dimensional tables;
- they prevent easy evolution of the application due to the non-separation between data management models, form and report definition. All these tasks must be accomplished by the user within a grid of cells, a cell being able to contain a piece of data, a formula and be used for presentation."

They conclude that the consequence of these limitations is that spreadsheets may be useful for simple personal applications but are very risky for more complex or institutionalised applications. It is these complex problems that exist in a socio-political business setting that are of interest to us. In the next section we introduce interpretation as a significant factor in the complex decision environment of modern business.

## 3. INTERPRETATION AS A COMPLICATING FACTOR

An individual can be seen as having a socially constructed, or nominalist, view of the world (Hirschheim and Klein 1989) and it is this social construction with its particular set of 'values, outlook, the worldview, (Weltanschauung), which makes a particular model meaningful' (Checkland and Holwell 1988). Without an appreciation of the worldview that the spreadsheet models were generated within it is unlikely that the full tacit connections will be obtained from the explicit content of the spreadsheet.

The worldview of the spreadsheet developer may also have been intrinsically flawed or may have been incorrectly articulated, captured or recorded. An individual 'expert' taking a defined role within an organisation may be acting with a distorted view of the world, but if there are no discernable dissonances then 'the performer can be fully taken in by his own act; he can be sincerely convinced that the impression of reality which he stages is the real reality' (Goffman 1978). If it is a respected role within the organisation, for example that of an 'expert', or experienced or senior member of the organisation then Goffman suggests that 'the audience is also convinced in this way about the show he puts on … then for the moment at least, only the sociologist or the socially disgruntled will have any doubts about the 'realness' of what is presented'. If the 'expert' is responsible for recording their knowledge into the spreadsheet then there is also the problem of actors presenting to the system only the polished and packaged end product of their reasoning. The errors and mistakes made during the reasoning process will have been corrected such that the 'telltale signs that errors have been made and corrected are themselves concealed … In this way an impression of infallibility, so important in many presentations, is maintained' (Goffman 1978)

If knowledge, then, is taken as 'dynamic human processing justifying personal belief toward the "truth" (Takeuchi and Nonaka in (Morey et al 2000)) developers must view the output of a spreadsheet as 'justified personal belief' rather than 'truth'. If an individual believes something to be true and acts as if it were true and as a result of that action perceives confirmatory outcomes, then they will continue to believe their perception to be real. Even if the spreadsheet output is incorrect there may be political or social reasons to project the outcome as if it were correct and this aspect of the process needs to be recognised as a contributory factor in the over analysis of the failure of spreadsheets. Kerzner (1998) notes that one implementation problem experienced in reporting project status or progress is that 'upper level personnel generally prefer the more traditional methods [of non-electronic reporting supported by trial and error], or simply refuse to look at reality because of politics'. He suggests that this political agenda may lead to data submitted to the board being based upon 'an eye-pleasing approach for quick acceptance, rather than reality'.

Panko and Halverson (1996) note that 'developers may even include deliberately incorrect data, or at least data from estimates that are dubious but support their cases' and that there is a natural tendency to select assumptions that fit individual expectations and desires. This indicates that even with a quantitative decision support tool the users are engaging in qualitative or interpretive actions and that these actions can significantly distort decision processes that are based on these actions and potentially lead to poor, ineffective or possibly dangerous business decision-making.

# 4. HELPING STUDENTS DEAL WITH COMPLEX AND INTERPRETIVE DEVELOPMENT

The operation of software tools is quite clearly a training specific activity, whereas the business problem solving approach would seem to fall more comfortably into the domain of education in the broader sense. Business computing has been in the curriculum of higher education for some considerable time, so it is rather surprising that the results of this exposure are not as successful as would be hoped for. Students do need software-specific skills but these must be matched with an educational process that helps them to contextualise those skills within a socio-technical business setting that is characterised by complexity, change and political action.

The difficulty that typically faces higher education is the need to set problems that are pre-digested and have predictable outcomes, ie a 'correct' answer against which a students performance can be measured. There is also often no effective way to simulate true business problems whose features often include perceived complexity, ambiguity, time and political pressure and so on. Such aspects of the real world are difficult to model within educational systems which are bounded by time and assessment implications, but our own work in this area suggests that considerable benefits can accrue if students are given problems to which there is no single correct outcome and where the exercise alerts them to the fact that in some cases a spreadsheet can only offer a guide rather than an accurate answer. Practical difficulties of assessment do arise, as the focus of activity becomes more process than product centred, and this becomes more difficult to measure in an objective fashion, but we argue that the benefits should outweigh these considerations. Greater concentration on the development of problem solving skills rather than offering a route to 'quick and dirty' implementation of an obvious answer to a problem should help students to become more careful in their adoption of simplistic, and potentially flawed, models, which are later implemented on a computer system.

## 4.1 Examples of the problems used and results obtained

**Balls in boxes – overly simplified or impossibly complex attempts to solve**

The basic problem here was based around the idea of packing balls in boxes, eg how many balls of a specific diameter could be packed in a box of given dimensions. Variations of this problem used mixtures of boxes and drums for packing and were presented as an industrial pellet production and packing process. Solutions developed ranged from simple volume division (volume of box divided by volume of ball multiplied by number of balls) through to attempts that recognised complex matrix packing variations for different sizes of ball. Perhaps the most creative solution was one that produced, for one specific combination, an answer that identified the optimum packing container as a drum, but pointed out that from a wider shipping perspective boxes would be more efficient. Some students were embarrassed to admit that they used marbles and jam-jars filled with water to explore the problem – they felt that this was quite a childish approach even though we praised them for their sensible experimentation. (Local stores did complain that they were selling out of marbles, even though this game was 'out of season'!) The solutions thus represented a spectrum from overly simplistic (the volume division approach) through to one that was too complex to model (the matrix model). Students found it difficult to accept that there was no 'perfect' answer to the question, tending to believe that what was required was further refinement of their solution rather than re-examination of the problem itself. These outcomes echoed previous work carried out by one of the authors with adult learners who were learning to develop computer programs in BASIC (Banks, 1988).

**Machine selection – interpretation of data**

We have also used a complex 'machine selection' problem in which students were required to develop, as groups, a spreadsheet to support a group of decision makers who were selecting one of five industrial production machines. With the same given information and specifications, of 100 groups of students 38 opted for machine A, 14 for machine B, and sixteen each for the remaining three machines. The spreadsheets contained few technical errors, in terms of errors that would cause the spreadsheet to malfunction, with the major differences in outcomes being a result of the interpretations that the students had applied to the data provided. The range of problems that were evident in the solutions for this problem included many of those mentioned in broader spreadsheet literature, including: poor design and layout, overly complex solutions that would be difficult for a group of users to understand, not comparing like with like, missing items in comparisons, assumed a 'right' outcome, little on-screen guidance for users, input and output data too far apart, little use of graphing, too much data on screen, incorrect filename protocol, ranking and weighting applied, and a good concept, but poorly executed.

**4.2 The student approach**

In addition to the mechanical problems raised earlier the following issues have been noted:
- that students tend to develop the outlines quickly without any real planning or problem solving.
- they are unable to filter out the data that is not required to solve the problem, attempting to incorporate all data provided. In particular, students are reluctant to remove data that has been provided for them.
    - students do not appear to recognise the importance of the interpretive act in both the design and the assessment of the validity of the output, believing their solution to be correct.
- students spend more time on presentation than testing the output, and apply fashion colours rather than good working practices.

However, in terms of attitude to this type of problem solving, many students who have been reticent in the early stages of the project, have risen to the challenge and developed considerably in terms of confidence in tackling unfamiliar and more complex problems.

**4.3 Problems of teaching**

In attempting to expose students to real world, complex problems we introduce risk, not only for the students, but for ourselves and our teaching team. It has not been uncommon for some support teaching staff to seek 'the answer' to the problem and to find it difficult to see that there are a number of answers based on the interpretive approach adopted by the students. We suspect that some tutors have found it embarrassing to admit that they do not know 'the' answer and have therefore tended to provide students with their own interpretations of the situation, these being accepted by the students as 'true' solutions. We have found it relatively easy to design simple spreadsheet problems but much more difficult to develop complex problems that stretch the students but can also be contained within a relatively short teaching semester.

## 4.4 How can we improve this situation?

Our aim is to help students, who will enter the business world as both constructors of spreadsheets and as consumers of the output from other developers' spreadsheets, to acquire a view of spreadsheets that acknowledges their value but also recognises the potential problems. For students who will become consumers of spreadsheets an option would be to provide them with constructed spreadsheets that produce different outputs for the same problem and ask them to evaluate the different scenarios.

For constructors of spreadsheets we must consider alternative approaches to developing skills in the design and the mechanical development of spreadsheets. We must also explore ways to emphasise the importance of documentation, testing and auditing of spreadsheets, particularly larger models; recognition that the models built are a product of the developers worldview rather than a 'scientific' or 'rational' process; recognition that the data present in spreadsheet and the way it is processed may not truly reflect the 'real' data as a result of deliberate or accidental misrepresentation of that data; and recognition that the data produced from spreadsheets is interpreted within the context of a broader socio-political business environment where even accurate output may be deliberately distorted.

This is a difficult task to embark upon but all of these factors must be taken into account when trying to understand why spreadsheets are causing such concern and potential damage in the business community. Computer programming has already been through much of this reflective thinking and spreadsheet developers, who carry out similar tasks to programmers in the development of a spreadsheet model, can learn from their conclusions. Dick (1985) offers a useful parallel in the need for a change of educational emphasis in his discussion of education for engineers: 'While engineering education has done a good job of teaching engineers how to use the tools of analysis, it has provided little or no training in the mechanics of forming and manipulating concepts. This has led many engineers to settle for mediocre results, rather than strive for more elegant, path-breaking solutions.' He suggests that the difference between inspired and mundane performance is not academic, it is more the application of both detailed analysis and conceptualisation to a problem to generate a creative solution and he states that: 'An engineer who consciously applies conceptualisation skills is better at defining and assessing problems, developing alternative solutions, and determining the best solutions. These abilities, in turn, are the basis for such breakthroughs as using old products in new ways or applying new technologies to revolutionize existing designs'

Another critic of the rigid thinking patterns produced by some engineering training and education is Sargent (1990) who comments that: 'Conventional engineering sciences and mathematics are oriented to understanding the physical world in which engineering occurs, and at applying that knowledge in techniques of analysis. Also many mathematical representations of the world are taught in which the "best" design is implicit, and can be determined by solving the relevant equations. Sargent (1990) continues: 'However, where the essence of the problem does not have a convenient representation, then degree courses are very poor in teaching methods of synthesis, of generating good potential designs which can then be analysed by more formal methods.' He argues that, in the real world, proposed designs '...will involve a variety of problems which cannot all be represented in the same mathematical description, and which may involve currently intractable problems which can be solved only approximately.' For such circumstances he indicates that the systematic generation of design alternatives and the selection of better ones based on incomplete information offers a productive path. The application of such approaches by end users in the computing world should help avoid some of the basic design problems that clearly plague even such apparently simple applications as spreadsheets.

## 5. CONCLUSION

Spreadsheets appear to be a relatively intuitive tool that are perceived to be deceptively easy to use and although extensively used by organisations have often proved to be misleading at best and dangerous at worst. The spreadsheet itself has some limitations in its ability to act as a vehicle for the development of complex business models and users may be relatively naïve in the development of models. Their basis for developing the model may be based on a unique worldview, or contain deliberate or unintended biases. The model may have been deliberately tampered with to generate acceptable figures for a specific instance to preserve a perceived 'expert' label for the developer.

Unless we can help students recognise these broader interpretive components of the process as well as use good mechanical design principles the problems of poorly developed and applied spreadsheets will not be alleviated in the near future. Spreadsheets are clearly useful to business but unquestioning reliance on their output is clearly dangerous given the possible obstacles that lie in the way of the development of a reliable output. We need to help future developers appreciate all of the literal and interpretive problems, social, political and technical, that influence the development of spreadsheets and to avoid the trap of developing applications that contain disasters waiting to happen. Above all we must avoid the mentality of the developer who left the comments below embedded in a large database application that ran successfully for one year after his efforts but then failed totally:

```
*apologies to anyone trying to maintain this code
*I just coded it and debugged it by running, and then
*added further pieces into it as they were needed.
*Anyway, I reckon that the more confusing it looks
*the better it must be!!!!!!!!!!
```